# Nanoscale optical switching of photochromic material by ultraviolet and visible plasmon nanofocusing

Takayuki Umakoshi[1,2,3,#,*], Hiroshi Arata[1,#], and Prabhat Verma[1]

1. Department of Applied Physics, Osaka University, Suita, Osaka 565-0871, Japan

2. Institute for Advanced Co-Creation Studies, Osaka University, Suita, Osaka 565-0871, Japan

3. PRESTO, Japan Science and Technology Agency, Kawaguchi, Saitama 332- 0012, Japan

#contributed equally

*Email: umakoshi@ap.eng.osaka-u.ac.jp

**ABSTRACT**

Optical control of electronic properties is essential for future electric devices[1,2]. Manipulating such properties has been limited to the microscale in spatial volume due to the wave nature of light; however, scaling down the volume is in extremely high demand. In this study, we demonstrate optical switching within a nanometric spatial volume in an organic electric material. Photochromic materials such as diarylethene derivatives exhibit semiconducting and insulating properties on ultraviolet (UV) and visible light, respectively, which are promising for optical switching and memory[3,4]. To control the wavelength between visible and UV light at the nanoscale, we employed plasmon nanofocusing, which allows the creation of a nanolight source at the apex of a metallic tapered structure over a broad frequency range by focusing of propagating plasmons[5,6]. We utilized an aluminum tapered structure and realized *in-situ* wavelength control between visible and UV light at the nanoscale. Using this method, nanoscale optical switching between the two states of diarylethene was demonstrated. The switching performance was confirmed for at least nine cycles without degradation. This demonstration would make a significant step forward toward next-generation nanoscale optoelectronic devices and stimulate diverse scientific fields owing to the unique concept of broadband plasmon nanofocusing.

Organic materials have played a vital role in the past few decades, contributing to drastic developments in flexible electronic technologies and devices, such as electronic memories, displays, and solar cells. The unique and multifunctional nature of organic electronic materials promises future applications that cannot be developed using silicon-based technologies[1,7–12]. Organic photochromic materials such as diarylethene derivatives, for instance, have garnered particular attention because of state control by light between semiconductors and insulators, which gives interesting functions to electronic devices, and thus holds significant potential for next-generation optical switching and memory[3,4,13–18]. Multilevel optical memory has recently been realized by employing photochromic materials[10,19], which has further stimulated related fields to accelerate its development. For such optoelectronic devices, scaling down the photochromic reaction volume has always been in exceptionally high demand for miniaturizing future optoelectronic devices, and is also an interesting subject for providing a new direction for fundamental chemistry and organic electronics. Meanwhile, it has been a long-standing challenge because of the diffraction limit of light, which does not allow light to be focused to less than approximately half the wavelength of light. Only under specific conditions, such as on an isolated single photochromic fluorescent molecule, the optical switching has been demonstrated[20,21].

In general, shrinking the illumination area of light is enabled by localizing light near a metallic nanostructure, which works as an optical antenna for light confinement through localized plasmon resonance[22,23]. It has been widely exploited as a popular tool for creating nanoscale reaction fields and inducing various photochemical reactions at the nanoscale with high efficiencies[24–29]. However, this is literally a resonant phenomenon that occurs within a narrow wavelength range[22,23,30]. It is useful for some photochemical reactions that require only a single light wavelength by tuning the resonance wavelength to a target wavelength. However, it is not a suitable method for photochemical events that involve multiple wavelengths, such as, optical switching of photochromic materials, because it requires visible and ultraviolet (UV) light for switching between two states[3,4]. *In-situ* control of the light wavelength is still highly challenging at the nanoscale. This fact has strongly restricted photochromic reactions at the nanoscale in a reversible manner for a few decades. Aperture probes have also been used as a near-field optical tool for photochromic reactions in the past few decades[31,32], however switching between two states at the nanoscale has not been demonstrated mostly due to a low throughput of UV light and several technical difficulties. Also, the light confinement volume has typically been limited to approximately one hundred nanometers, not truly down to the nanoscale.

In this study, we achieved optical switching within a nanometric volume of photochromic materials. To this end, we completely abandoned the method to create nanoscale light field from the optical antenna technique based on localized plasmon resonance. In this study, we adopted plasmon nanofocusing as an alternative method for inducing a nanoscale light source. Plasmon nanofocusing is a phenomenon in which plasmons are excited near the broader end of a metallic tapered structure and propagate toward the apex with adiabatically compressing their energies, which eventually create a strong light field localized at the nanometrically sharp apex of the tapered structure[33–37]. Because a nanoscale light source is excited through the propagation of plasmons, rather than the localized plasmon resonance, we have demonstrated in our previous studies that it works over an extremely wide frequency range from the entire visible to near-infrared regions[5,6]. We have proposed the generation of a white nanolight source through broadband plasmon nanofocusing and successfully exploited it for super-resolution hyperspectral optical imaging[5], which has exhibited great potential for broadband plasmon nanofocusing not only for optical imaging but also for diverse scientific fields as a novel tiny light source[38–40]. Therefore, we adopted broadband plasmon nanofocusing for nanoscale optical switching in this study. Figure 1 illustrates an overview of the nanoscale optical switching that we demonstrated using plasmon nanofocusing. We developed a totally different system for plasmon nanofocusing in this study, while we used it on a cantilever tip in the previous study. We fabricated a metallic tapered structure on a substrate, and further fabricated a protective layer and a hole to realize optical switching at a nanometrically sharp apex of the tapered structure, details of which are described later. We further expanded the wavelength range of plasmon nanofocusing even down to the UV region for the optical switching. To this end, we employed aluminum as a plasmonic material suitable in the UV region. We prepared the aluminum tapered structure, and fabricated a slit structure as the plasmon coupler because the slit works for a wide wavelength range. We successfully controlled the wavelength of the nanolight source from the UV to visible regions. Plasmon nanofocusing provides another advantage, which is the background-free characteristic of unwanted noise caused by an excitation laser[35–37]. Because plasmons are excited by an excitation laser at the slit located far from the apex of the tapered structure, the nanolight source is spatially well separated from the excitation light, which is not the case for optical antennas because it requires directional illumination of the optical antennas. In particular, this is highly effective for optical switching because we can avoid unwanted photochromic reactions by the excitation light near the nanolight source. Considering the high freedom of the choice of wavelength and background-free features, we utilized plasmon nanofocusing as the best suitable way to achieve optical switching within a small volume.

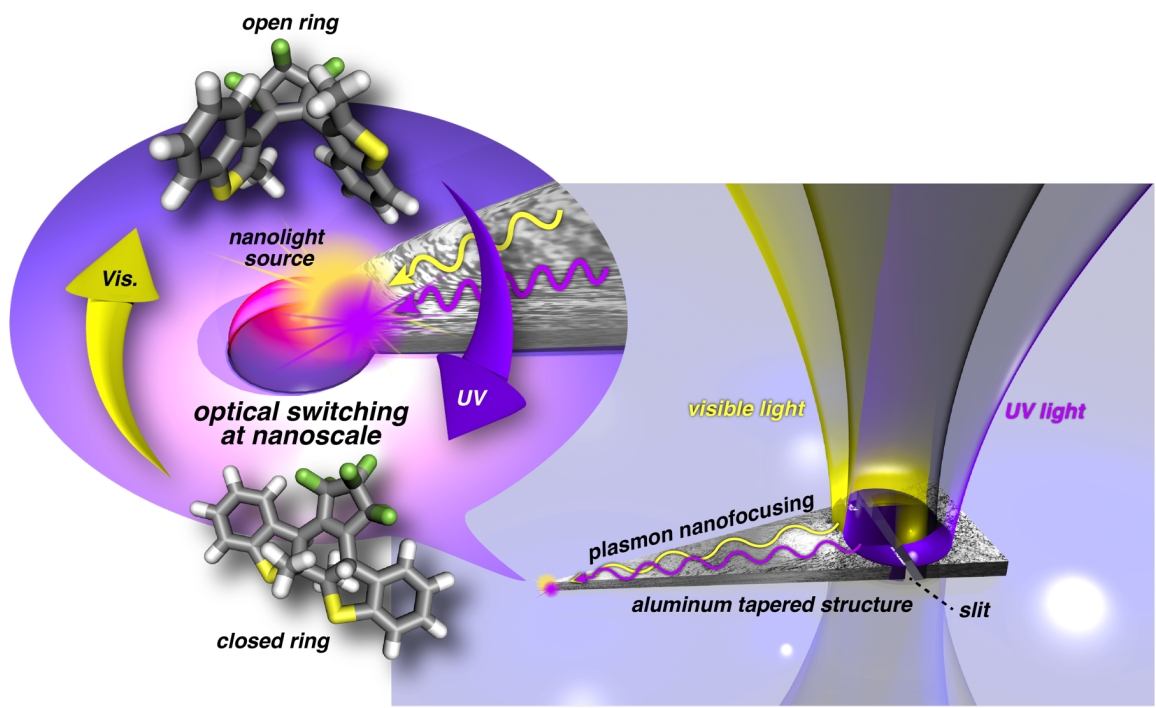

**Figure 1. Schematic of nanoscale optical switching of photochromic material by plasmon nanofocusing.** Nanoscale visible and UV light sources created at the apex of the aluminum tapered structure optically switch electric properties of photochromic material at the nanoscale.

**Photochromic materials for optical switching**

We selected diarylethene derivative among various types of photochromic materials, such as azobenzene and spiropyran, because it is one of the most popular and best-suited materials for device applications owing to its high thermal stability and excellent electrical properties[3,4]. The diarylethene molecule takes a closed-ring form that exhibits semiconducting properties when irradiated with UV light and is converted to an open-ring form when irradiated with visible light, which shows insulating properties. Figure 2 shows the absorption spectra of the open- and closed-ring forms of diarylethene molecules in which the molecules are dissolved in PMMA/anisole solution (see Methods and Fig. S1 for further detail). The absorption spectra of the open- and closed-ring forms of the sample solution were measured after visible and UV light irradiation, respectively. Upon visible light illumination of closed-ring diarylethene molecules, the broad absorption peak at approximately 530 nm disappeared while the absorbance increased in the UV range. This spectral change indicates a conversion from the closed-ring form to the open-ring form. In contrast, the absorption peak at 530 nm appeared upon UV light illumination to convert open-ring diarylethene molecules to closed-ring molecules. It was also visually evident as the color of the solution changed from transparent to red. We can thus understand through the absorption measurement which state the diarylethene molecules take between semiconductor and insulator.

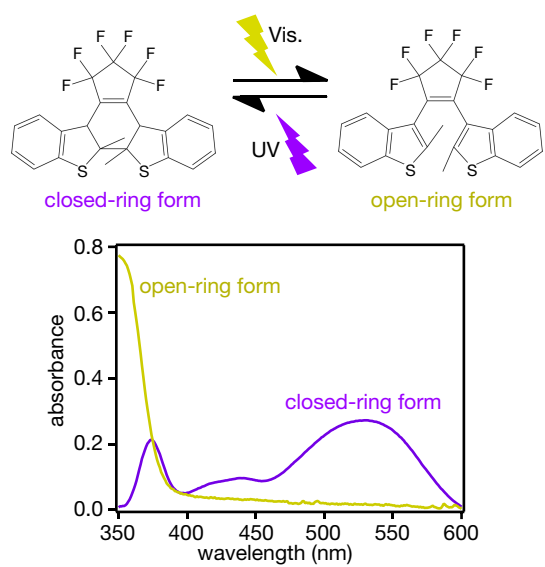

**Figure 2. Absorption spectra of diarylethene solution.** Absorption spectra of open-ring diarylethene molecules (yellow curve) and closed-ring diarylethene molecules (purple curve).

**Plasmon nanofocusing of visible and UV light for optical switching**

Plasmon nanofocusing can be excited over a wide frequency range; however, the excitation of plasmon nanofocusing in the UV range is not straightforward. Broadband plasmon nanofocusing has been demonstrated with a silver or gold tapered structure in previous studies, in which it covers the entire visible range to the entire near-infrared range[5,6]. However, plasmon nanofocusing is comparatively inefficient in the shorter wavelength region due to the much shorter propagation length of plasmons, and thus UV plasmon nanofocusing has not yet been demonstrated. Therefore, we selected aluminum instead of silver or gold to push the wavelength range down toward the UV range, as aluminum exhibits great plasmonic properties even in the UV range[41,42]. We evaluated the optical properties of an aluminum tapered structure for UV plasmon nanofocusing via numerical simulation (See Methods and Fig. S2 for details). First, we used an incident light with wavelength of 530 nm for testing plasmon nanofocusing in the visible range, which corresponds to the absorption wavelength of a closed-ring diarylethene molecule. A distribution map of the electric field intensity was simulated around the apex of the tapered structure during the plasmon nanofocusing process (Fig. 3(a)). We confirmed that the light field was well confined within a nanometric volume at the apex through plasmon nanofocusing. The wavelength was thereafter changed to 375 nm on the same simulation model. 375 nm is an efficient wavelength for the conformational change of diarylethene molecules from an open-ring to a closed-ring form. As shown in Fig. 3(b), the UV plasmon nanofocusing was also observed at 375 nm. Although the electric field intensity was 3–4 times weaker than that for visible light because of the shorter plasmon propagation length, we still obtained sufficient intensity for practical use. We therefore concluded that aluminum is a highly promising material for UV plasmon nanofocusing. These results suggest that a nanolight source can be created in both wavelength ranges, and thus plasmon nanofocusing efficiently enables optical switching of diarylethene molecules in such a tiny volume. We also found that other metals, such as silver and gold, were not capable of exciting UV plasmon nanofocusing, as shown in Fig. S3.

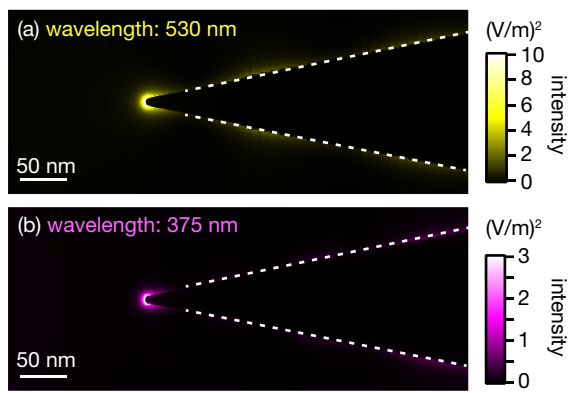

**Figure 3. Simulation of plasmon nanofocusing of visible and UV light on an aluminum tapered structure.** Distribution maps of electric field intensity near the apex of an aluminum tapered structure during plasmon nanofocusing. (a) wavelength: 530 nm. (b) wavelength: 375 nm.

It should be emphasized as another crucial property that plasmon nanofocusing is free from chromatic aberration, as the nanofocusing process is based on propagating plasmons bound on a tapered structure as a surface wave. It is usually difficult to focus both UV and visible light in a similar focal size at the same location using ordinary optics because of the chromatic aberration caused by the large wavelength difference between visible and UV light. This can result in a large difference in the photochromic reaction volume and a deteriorated switching performance. However, by nanofocusing light through a metallic tapered structure, both UV and visible light are focused exactly at the same location because the propagating plasmons automatically go to the apex. In addition, we confirmed in the simulation that the spatial extent of the localized light field for both UV and visible light was also the same at the nanoscale, which was approximately 12–13 nm (Fig. S4), because it was determined by the apex size rather than the wavelength itself[22,23]. This outstanding property is highly effective, particularly for switching applications because we can reversibly excite photochromic reactions at exactly the same location and volume with nanometric precision.

After confirming both the UV and visible plasmon nanofocusing on an aluminum tapered structure for nanoscale optical switching in the numerical simulation, we went ahead to fabricate a tapered structure similar to the simulation model on a glass substrate as a switching platform, followed by the fabrication procedure shown in Figure 4(a). Aluminum tapered structures were fabricated by the physical vapor deposition of a thin and smooth aluminum layer and subsequent focused ion beam (FIB) lithography. Because propagating plasmons can cause photochromic reactions during propagation on the surface of a tapered structure before reaching the apex, we made a 100-nm-thick polymer (PMMA) coating by spin coating as a protective layer to prevent diarylethene molecules from directly touching the tapered structure. We further created a hole by FIB milling to expose the apex of the tapered structure such that diarylethene molecules could only be present near the apex of the tapered structure. The fabrication details are described in Methods and in Fig. S5. We also examined the surface roughness of the aluminum layer, which was less than a few nanometers and sufficiently smooth for plasmon propagations (Fig. S6.)[37]. Figure 4(b) shows the scanning electron microscopy (SEM) image of the fabricated structure. The tapered structure was not clearly visible because it was embedded in the polymer layer. In the enlarged SEM image of the hole, as shown in the inset of Fig. 4(b), we observed a tapered structure apex exposed from the polymer coating, as indicated by the white dotted circle. In Fig. 4(b), to clearly show the apex exposed from the polymer coating, we purposely etched a large portion of the apex by FIB milling such that the apex size was as large as ~50 nm. In practice, it is possible to maintain a much sharper apex by precisely adjusting the hole position in the FIB process.

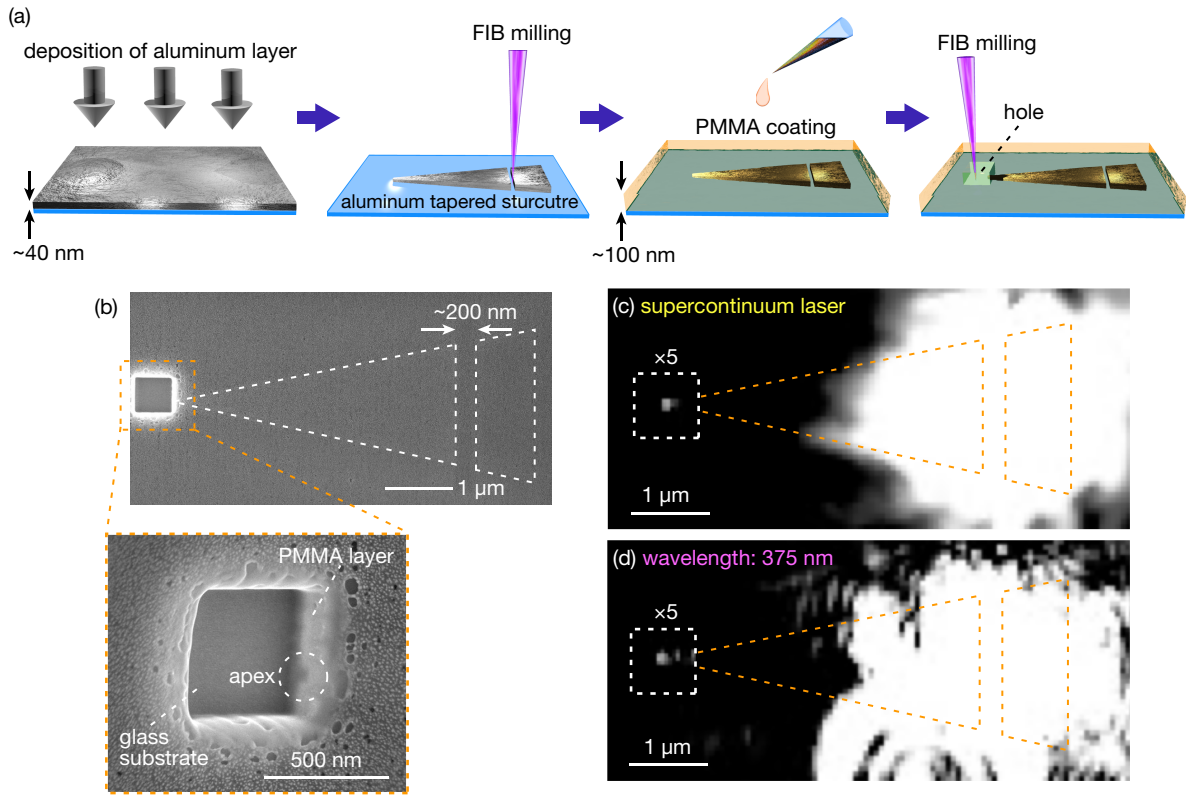

**Figure 4. Fabrication of the aluminum tapered structure and optical observation of plasmon nanofocusing.**
(a) Fabrication procedure of aluminum tapered structures. (b) SEM image of the fabricated tapered structure imbedded in a polymer film. The inset shows an enlarged SEM image near the apex, where the apex exposed from the polymer coating was observed in the hole. (c) Optical image of plasmon nanofocusing with super continuum laser. (d) Optical image of plasmon nanofocusing excited on the same tapered structure as (c) with the UV laser (wavelength: 375 nm). The image contrast was increased five times in the white dotted squares for visual clarity.

We implemented optical measurements with the fabricated structure to observe plasmon nanofocusing under both visible and UV light. Figure 4(c) shows an optical image of an aluminum tapered structure in which a supercontinuum laser that covers the entire visible range irradiates the slit. We observed a bright optical spot at the apex as a scattering signal from the nanolight source generated through plasmon nanofocusing, which was spatially well separated from the incident laser. We thereafter switched from a supercontinuum laser to a UV laser with a wavelength of 375 nm. A bright optical spot was clearly observed on the same tapered structure even with the UV laser (Fig. 4(d)). This confirms that our fabricated tapered structure is effective for both the UV and visible plasmon nanofocusing to implement nanoscale optical switching.

**Optical switching at nanoscale**

As we developed a switching system for diarylethene molecules, in which we can freely switch the wavelength of the nanolight source between the visible and UV range, we performed optical switching of diarylethene molecules in this setup. We dropped a 100 mM diarylethene solution on the tapered structure and allowed it to dry. First, we excited plasmon nanofocusing with a supercontinuum laser as visible light in this setup. All diarylethene molecules were converted to a closed-ring form by shining the entire substrate with a UV lamp prior to the open-ring reaction using visible plasmon nanofocusing. While broadband plasmon nanofocusing is utilized for optical switching, it is also useful for monitoring the optical properties of the diarylethene molecules at the apex. By detecting the broadband scattering spectrum from the apex, it is possible to evaluate changes in the absorption properties of the samples only near the apex[5] (See Methods and Fig. S7 for more details). We excited the photochromic reaction around the apex through plasmon nanofocusing with a supercontinuum laser while observing changes in the scattering spectra. A reference spectrum was also obtained prior to the experiment. The detected scattering spectra and the reference spectrum were used to obtain the absorption spectra. As shown in Fig. 5(a), we observed the conversion from the closed-ring form to the open-ring form, which was induced at the nanoscale only near the apex through the plasmon nanofocusing of the supercontinuum laser. The absorption peak at approximately 530 nm, which originated from closed-ring diarylethene molecules, gradually disappeared within 10 s. The flat absorption feature in the visible range corresponds to that of open-ring diarylethene molecules. This confirms that molecules within the nanofocused light were converted to the open-ring form in 10 s. We then illuminated the slit of the same tapered structure with the UV laser and observed the recovery of the absorption peak within 10 s, as shown in Fig. 5(b). Here, while we illuminated the UV laser for the photochromic reaction, the supercontinuum laser was also used for a short duration to obtain the scattering spectra. However, this short exposure did not make any noticeable change in absorption spectra. Figures 5(a) and (b) clearly indicate that the closed-ring diarylethene molecules turned to the open-ring form and returned to the closed-ring form by visible and UV plasmon nanofocusing. The absorption spectra of the diarylethene solution, measured with an ordinary UV-Vis spectrometer, are shown in Fig. 5(c) as a reference, which corresponds well with the change in the absorption spectra obtained with the plasmon nanofocusing technique. As a control experiment, we conducted the same experiment without a hole at the apex (Fig. S8), indicating that the tapered structure was fully covered with the polymer layer, and diarylethene molecules could not exist near the apex of the tapered structure. In this situation, we did not observe any changes in the absorption spectra, suggesting that the observed absorption change was attributed to the photochromic reaction induced only at the apex. We confirmed that the two states of diarylethene were optically switched within a nanometric volume.

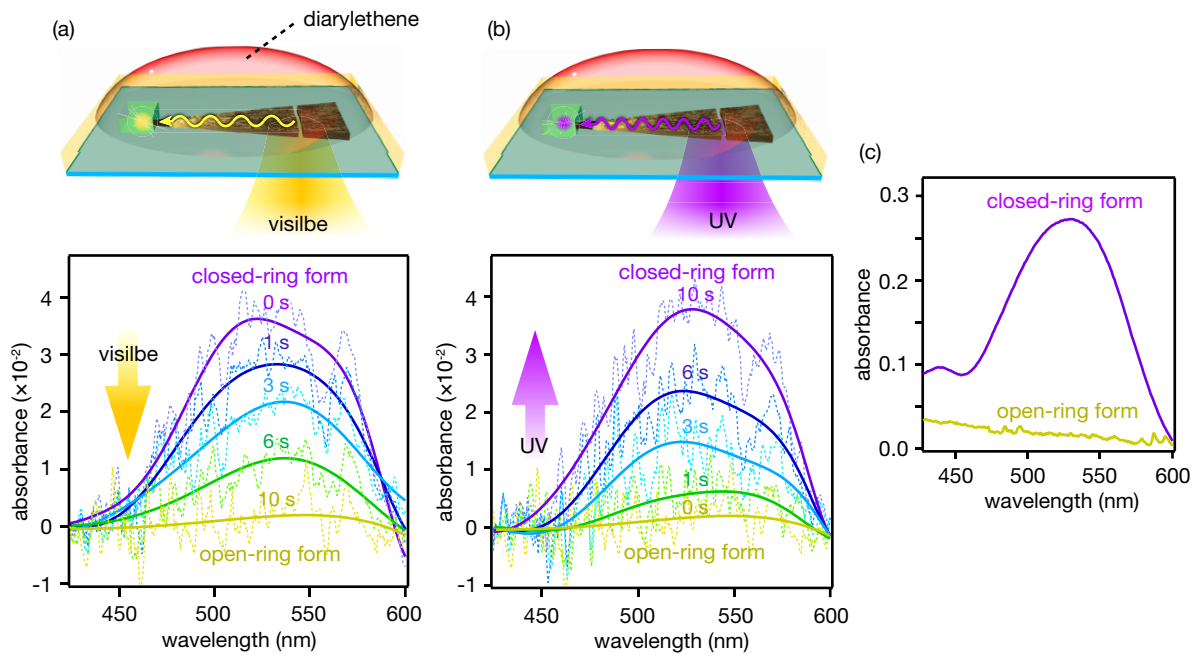

**Figure 5. Nanoscale photochromic reactions induced by plasmon nanofocusing.** (a) Time variation of absorption spectra by plasmon nanofocusing of visible light to turn diarylethene molecules from closed-ring form to open-ring form. (b) Time variation of absorption spectra by plasmon nanofocusing of UV light to turn back to closed-ring form. (c) Absorption spectra of diarylethene solution as a reference.

      Furthermore, we repeated this switching cycle to evaluate the performance of nanoscale optical switching. As shown in Fig. 6(a), we confirmed that the closed-ring form and open-ring form were switchable multiple times by alternate visible and UV plasmon nanofocusing. We observed it up to nine cycles without any degradation in the absorbance amplitude (Fig. 6(b)). Here, the absorbance change at 530 nm was plotted for each cycle. All the absorption spectra are shown in Fig. S9. We successfully demonstrated nanoscale optical switching of photochromic materials by employing a novel broadband plasmon nanofocusing.

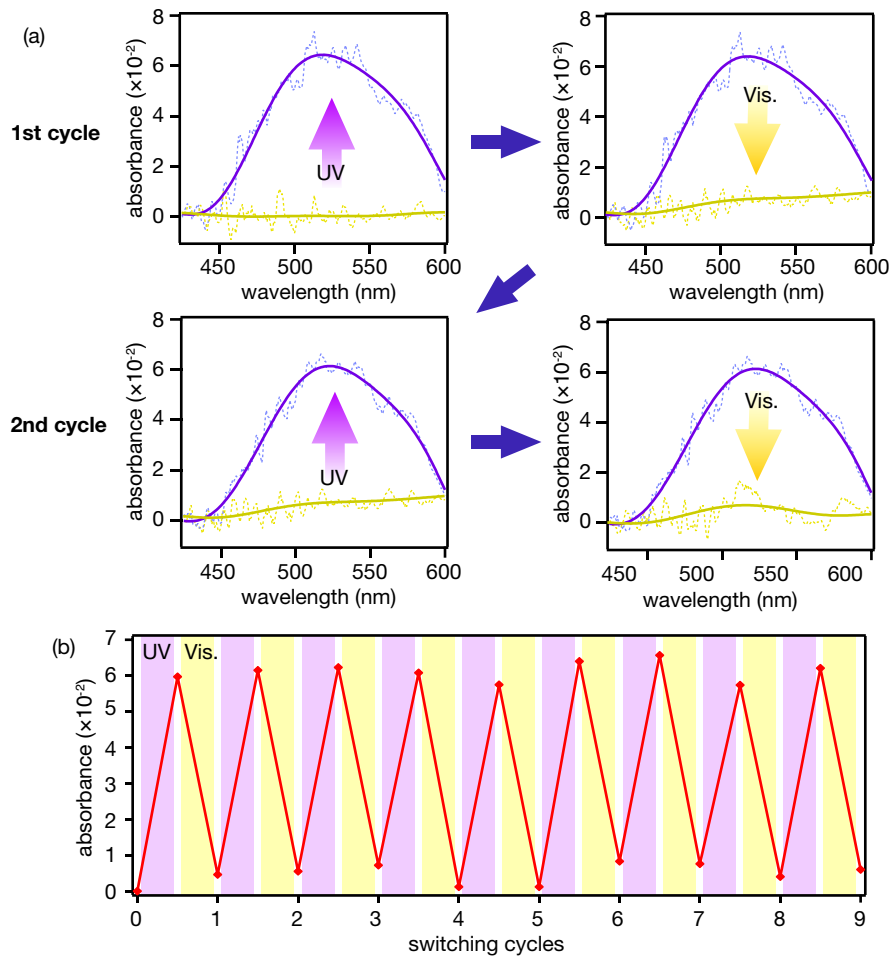

**Figure 6. Nanoscale optical switching between open- and closed-ring forms of diarylethene molecules.** (a) Absorption spectral change of diarylethene molecules during the nanoscale optical switching. (b) Repeatability of the nanoscale optical switching. Absorbance change at the wavelength of 530 nm was plotted.

**Conclusion and Discussion**

Nanoscale optical switching of organic photochromic materials was achieved by exploiting the unique advantages of plasmon nanofocusing. As organic photochromic materials such as diarylethene molecules are indispensable materials for future organic optoelectronic devices, this demonstration would be a significant step forward for employing photochromic materials in the nanoscale regime and for realizing nanoscale organic optoelectronic devices. It was not possible to evaluate the actual size of the nanoscale optical switching at this moment; however, it should be as small as the apex size (10–20 nm)[22,23]. An aperture probe would also be used as another tool to generate near-field visible and UV light[31,32]. However, the size is typically limited to ~100 nm due to limited throughput. In this sense, plasmon nanofocusing has a strong benefit in truly accessing the nanoscale. Although more developments will be necessary to exploit this technique for actual devices, we expect that further improvements will be made in the near future.

The scaling down of a photochemical phenomenon that we achieved using broadband plasmon nanofocusing would have a huge impact not only for organic electronics but also for various scientific fields. Our concept of broadband plasmon nanofocusing provides an innovative tool for photochemistry, as it can be applied to several different photochemical reactions. Different photochemical reactions can be induced at exactly the same location within a nanometric volume by applying different wavelengths sequentially, which significantly improves the flexibility of nanoscale photochemistry. We can provide a small reaction field where the light wavelength can be freely controlled from the UV to the near-infrared range for chemical reactions. Furthermore, the background-free scheme effectively works as another strong advantage, as no by-products are produced by incident light.

We have also made several developments in this research to realize nanoscale optical switching, and the first demonstration of UV plasmon nanofocusing is of great importance. This extremely expanded the capability of the plasmon nanofocusing technique. The wavelength range of plasmon nanofocusing has now been extended to such a high-energy wavelength region, which will call for several other potential applications not only in photochemistry but also in wide fields of optics and photonics. The broadband property of plasmon nanofocusing spanning over the UV to near-infrared regions would be highly beneficial for various optical analytical techniques such as absorption spectroscopy, Raman spectroscopy, and photoluminescence measurements. Moreover, further extending the wavelength range toward the mid-infrared region to create a mid-infrared broadband nanolight source, highly sensitive molecular sensing and analysis through infrared absorption spectroscopy would be possible. In this study, we have used broadband plasmon nanofocusing in a more active way to manipulate physical and chemical properties of samples. More precise and flexible manipulation of samples would be expected by involving nanoscale optical trapping via the broadband plasmon nanofocusing. Plasmon nanofocusing is becoming an innovative but practical tool for providing high freedom of wavelength to nanoscale experiments, which possesses great promise in diverse scientific fields. The nanoscale optical switching that we achieved in this study would be a milestone for future developments in fields ranging from organic electronics and photochemistry to nanophotonics and nanotechnology.

## METHODS

### Finite-difference time-domain (FDTD) simulation

We used the FDTD method (Poynting, Fujitsu Co. Ltd.) to simulate the electric field distributions around an aluminum tapered structure during the plasmon nanofocusing process. The apex size and angle of the tapered structure were 10 nm and 25 degrees, respectively. The thickness was set to 40 nm. The slit was designed as a plasmon coupler on the tapered structure. The slit width was 200 nm, and the distance between the apex and slit was 4 μm. The slit was illuminated with visible or UV light. The wavelength for visible light was 530 nm, and that for UV light was 375 nm, which are efficient wavelengths for open- and closed-ring reactions, respectively. The real and imaginary parts of permittivity of aluminum were set to -20.61 and 3.92, respectively, for the wavelength of 375 nm, and -40.07 and 11.92, respectively, for the wavelength of 530 nm. The tapered structure was placed on a glass substrate. The calculation model was discretized for simulation. The grid size was assigned non-uniformly. It was 1 nm near the apex, and in the range of 10–20 nm in the other areas for sufficient calculation accuracy and affordable simulation time. Further details are provided in the Supplementary Information (Fig. S2).

### Fabrication of aluminum tapered structures

First, a thin aluminum layer was deposited by physical vapor deposition (VPC-1100, ULVAC, Inc.) on a coverslip that was cleaned with piranha solution in advance. The deposition rate and thickness were 1.5 nm/s and 40 nm, respectively. A smooth aluminum layer was obtained under these conditions (Fig. S6). An aluminum tapered structure and slit structure were fabricated via FIB lithography (FB2200, Hitachi High-Tech, Corp.). The distance between the slit and the taper apex was 4 μm, which was sufficiently large to spatially separate the nanolight source at the apex and the incident light at the slit.

Next, we dropped 10 μL of PMMA solution (495PMMA A Resists, KAYAKU Advanced Materials, Inc.) on the substrate, and spin-casted it using a spin coater. The coating process started at a rotation rate of 500 rpm for 5 s, gradually increased from 500 to 1,000 rpm in the next 30 s, and continued further for the next 60 s at 1,000 rpm. The PMMA layer was formed on the tapered structure with a thickness of approximately 100 nm, which was confirmed by atomic force microscopy.

A hole near the apex of the tapered structure was subsequently fabricated by FIB milling. A thin platinum layer was deposited with a thickness of 1.5 nm as a conductive layer on the PMMA layer using an ion-sputtering device (JEC-3000FC, JEOL Ltd.) prior to the fabrication of the hole. A platinum coating was necessary for FIB observation and fabrication. The hole size was $500 \times 500$ nm$^2$, as shown in Fig. 4(b). The hole position was set in such a way that it just exposed the apex of the structure. We thereafter dropped 10 µL of 100 mM diarylethene solution on the substrate and allowed it to dry for 15 min for the optical switching experiments of diarylethene molecules. The fabrication details are explained in the Supplementary Information (Fig. S5).

**Experimental setup for optical switching**

A supercontinuum laser (SuperK COMPACT, NKT Photonics) and a UV laser (LSR375NL-150, NaKu Technology Co., Ltd.) were used as the visible and UV light sources, respectively. The better coherence of the supercontinuum laser compared with other ordinary white light sources allows to tightly focus visible light to the slit. They were focused on the slit using an oil-immersion objective lens (NA: 1.45, UPLXAPO100XO, Olympus). The polarization of the incident light was oriented perpendicular to the slit, where the coupling efficiency between the incident light and plasmons was maximized. Optical images of the plasmon nanofocusing were obtained using a CMOS camera (VLG2020e-SP, Visualix). To evaluate changes in the absorption properties of diarylethene molecules at the apex of the tapered structure, the scattered signal from the apex through plasmon nanofocusing of visible light was detected by a CCD camera (PIXIS100BR_eXcelon, Teledyne Technologies Inc.) through a spectroscope (IsoPlane 160, Teledyne Technologies Inc.). A pinhole (diameter: 200 µm) was used as the confocal element to extract only the scattered signal from the apex. The details are described in the Supplementary Information (Fig. S7).

**ACKNOWLEDGEMENTS**

This research was partly supported by JSPS Grant-in-Aid for Scientific Research (A) 19H00870, JSPS Grant-in-Aid for Scientific Research (B) 20H02658, JST PRESTO Grant Number JPMJPR19G2, SECOM Science and Technology Foundation, Toyota Physical and Chemical Research Institute, and Research Foundation for Opto-Science and Technology.

**AUTHOR CONTRIBUTIONS**

T.U. conceived and designed this project, and wrote the manuscript. H.A. performed the experiments. T.U. and P.V. supervised this research. All authors contributed discussion of the results and commented on the manuscript.

**COMPLETING INTERESTS**

The authors declare no completing interests.

*Supplementary Information*

# Nanoscale optical switching of photochromic material by ultraviolet and visible plasmon nanofocusing

Takayuki Umakoshi[1,2,3,#,]*, Hiroshi Arata[1,#], and Prabhat Verma[1]

1. Department of Applied Physics, Osaka University, Suita, Osaka 565-0871, Japan

2. Institute for Advanced Co-Creation Studies, Osaka University, Suita, Osaka 565-0871, Japan

3. PRESTO, Japan Science and Technology Agency, Kawaguchi, Saitama 332- 0012, Japan

#contributed equally

*Email: umakoshi@ap.eng.osaka-u.ac.jp

# Contents

## 1. Ultraviolet-visible (UV-Vis) absorption measurements of diarylethene solution

We used 1,2-bis[2-methylbenzo[*b*]thiophen-3-yl]-3,3,4,4,5,5-hexafluoro-1-cyclopentene (B5623, Tokyo Chemical Industry, Co. Ltd.) as one of the most typical diarylethene derivatives. The molecular structure is shown in Fig. S1. Diarylethene powder was dissolved in PMMA/anisole solution (495PMMA A Resists, KAYAKU Advanced Materials, Inc.). The solution was stirred for a few minutes using a vortex mixer. The concentration of diarylethene used was 100 mM. Figure S1 shows pictures of the diarylethene solution in either closed-ring or open-ring form. It is transparent under visible light irradiation. It becomes reddish under ultraviolet (UV) light irradiation and turns back to transparent by irradiating with visible light reversibly, as seen in the pictures. To measure the absorption spectra, the solution was diluted to a concentration of 20 mM. We used a UV-Vis spectrometer (UV-3600, Shimadzu Corporation) to investigate the absorption properties of diarylethene. To measure the absorption spectrum of the closed-ring form of diarylethene, we illuminated the diarylethene solution with a UV lamp for a long time immediately before absorption measurement. Similarly, we illuminated the solution with visible light before measuring the absorption spectrum of the open-ring form of diarylethenes.

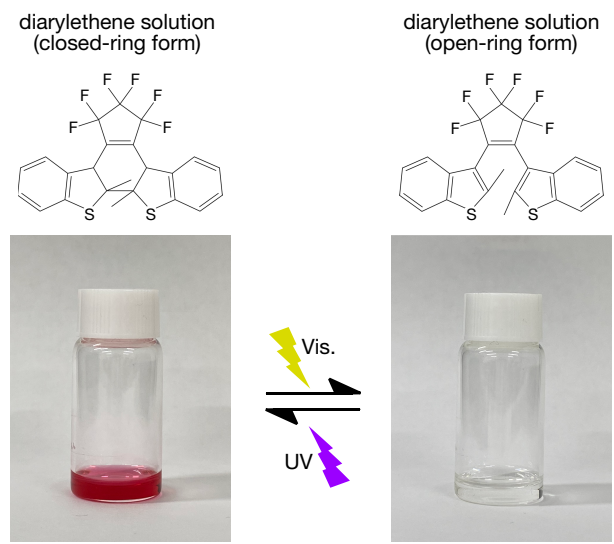

**Figure S1.** Pictures of diarylethene solution after visible and UV light irradiation.

## 2. Numerical simulation model

We used the finite-difference time-domain (FDTD) method (Poynting, Fujitsu Co. Ltd.) to calculate the optical properties of the tapered aluminum structure on plasmon nanofocusing. We designed a tapered structure as shown in Fig. S2. The apex size, angle, and thickness of the tapered structure were 10 nm, 25 degrees, and 40 nm, respectively, as described in our previous study[1]. The slit was located as a plasmon coupler with a width of 200 nm. The distance between the slit and apex was 4 μm to ensure sufficient separation between the incident light and the nanolight source generated through plasmon nanofocusing. The slit was illuminated with incident light of wavelength 375 nm or 530 nm. The real and imaginary parts of permittivity of aluminum were set to -20.61 and 3.92 for the wavelength of 375 nm, and -40.07 and 11.92 for the wavelength of 530 nm. The tapered structure was placed on a glass substrate. The simulation model was discretized for the numerical simulation. The grid size was non-uniformly assigned, where it was 1 nm near the apex, and 10–20 nm in the other areas for sufficient calculation accuracy and affordable simulation time.

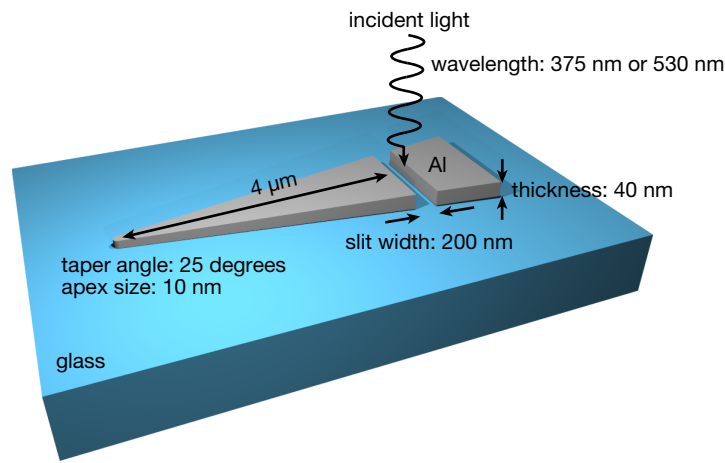

**Figure S2.** Schematic of the FDTD simulation model for plasmon nanofocusing

## 3. Plasmon nanofocusing with silver and gold tapered structures

As a comparison to the aluminum tapered structure, we simulated distribution maps of the electric field intensity near the apexes of silver and gold tapered structures. Silver and gold are representative plasmonic materials. The structural designs were exactly the same as those in Fig. S2. The real and imaginary parts of the permittivity of silver were -2.75 and 0.67 for the wavelength of 375 nm, and -10.55 and 0.84 for the wavelength of 530 nm. These values for gold are -0.76 and 6.47 for the wavelength of 375 nm, and -6.29 and 2.04 for the wavelength of 530 nm. In the case of silver, plasmon nanofocusing was observed at 530 nm (Fig. S3(a)), but not at 375 nm (Fig. S3(b)). This is even worse in the case of gold because plasmon nanofocusing was not observed at wavelengths of 530 and 375 nm (Figs. S3(c) and S3(d)), which corresponds to the results of our previous study[1]. Gold exhibits better optical properties for plasmon nanofocusing in the near-infrared region. These results further suggest that aluminum is suitable for simultaneously exciting both the UV and visible plasmon nanofocusing.

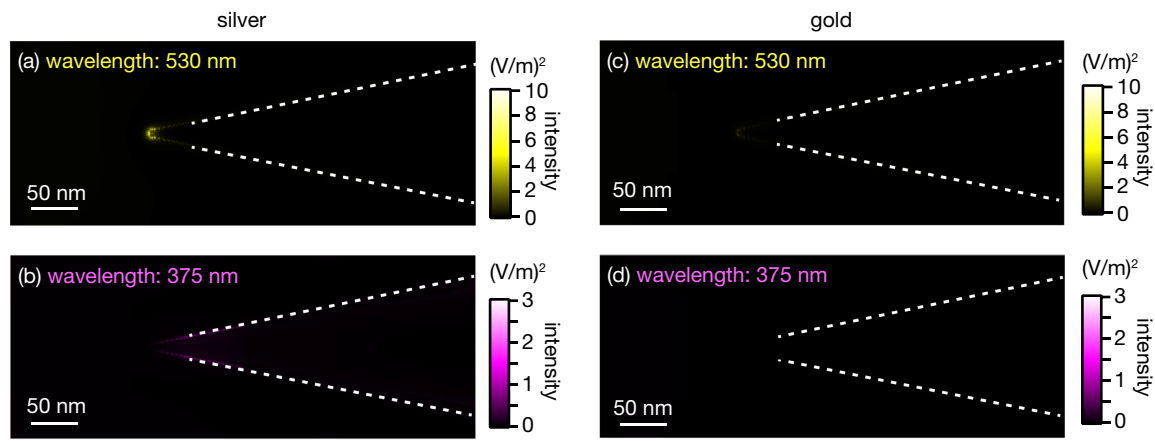

**Figure S3.** Distribution maps of electric field intensity near the apexes of silver and gold tapered structures during plasmon nanofocusing. (a) Silver tapered structure at the wavelength 530 nm. (b) Silver tapered structure at the wavelength 375 nm. (c) Gold tapered structure at the wavelength 530 nm (d) Gold tapered structure at the wavelength 375 nm.

## 4. Spatial extent of the localized light field generated through UV and visible plasmon nanofocusing

We investigated the size of the localized light field generated by plasmon nanofocusing using FDTD simulations. We took the line profiles of the electric field intensity along the red dotted lines in the right panels of Figs. S4(a) and S4(b), which are the same images as Fig. 3 in the main text. The red dotted lines are located 6 nm from the apex of the tapered structure to avoid any artifacts from the discretization of the simulation model. The line profile exhibited a Gaussian distribution. The wavelength is 530 nm in Fig. S4(a). The full width at half maximum (FWHM) of the Gaussian peak was approximately 13 nm to evaluate the spatial extent of the localized light field, which was almost comparable to the apex size. We obtained a similar line profile at a wavelength of 375 nm (Fig. S4(b)). The FWHM for a wavelength of 375 nm is estimated to be approximately 12 nm. Therefore, despite the large difference in wavelengths, incident light at both wavelengths is nano-focused in almost the same volume because it is determined by the apex size rather than the wavelength itself in plasmon nanofocusing[2]. This chromatic-aberration-free characteristic is highly effective in achieving closed- and open-ring reactions within exactly the same nanometric volume.

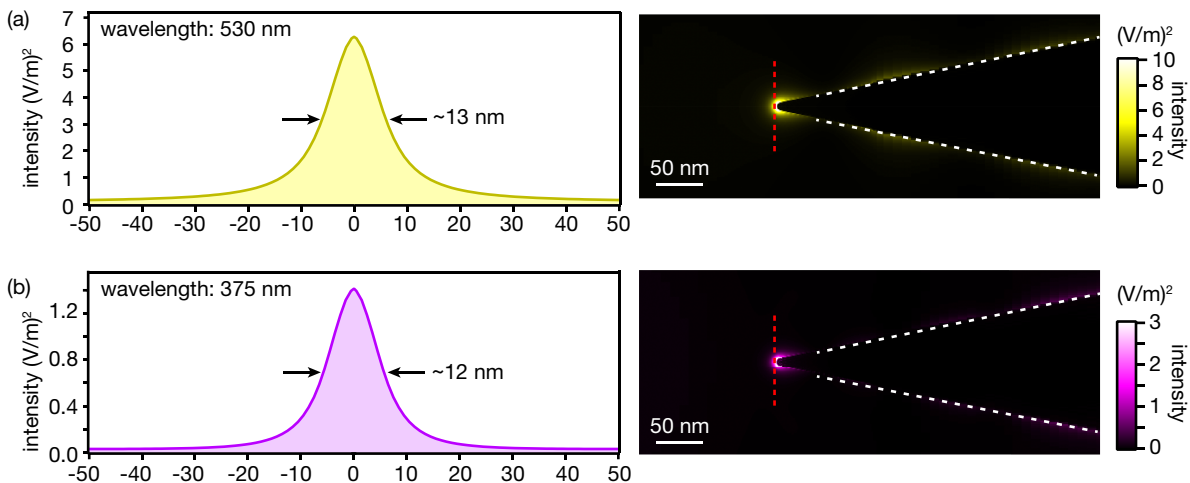

**Figure S4.** Spatial extent of the localized light field at the apex of the aluminum tapered structure during plasmon nanofocusing at wavelengths of (a) 530 and (b) 375 nm. We evaluated line-profiles of electric field intensity at the apex. Line-profiles were obtained from the red dotted lines on the right panels, showing the distribution maps of electric field intensity around the apex, which are same as Fig. 3 in the main text.

**5. Fabrication procedure of aluminum tapered structures for the nanoscale optical switching**

Herein, we describe the details of the fabrication procedure of the aluminum tapered structures used for nanoscale optical switching. A coverslip was cleaned in piranha solution for 30 min at 120 °C. A thin aluminum layer was thereafter deposited on the coverslip via vacuum vapor deposition (VPC-1100, ULVAC, Inc.). The evaporation rate was as fast as 1.5 nm/s such that a smooth aluminum layer was formed, as discussed in the next section. The thickness of the aluminum layer is 40 nm. We then fabricated tapered structures using focused ion beam (FIB) lithography (FB2200, Hitachi High-Tech, Corp.), as shown in Fig. S5(a). The distance between the slit and apex was ~4 μm. Next, the fabricated tapered structure was embedded in a polymer layer (PMMA) (495PMMA A Resists, KAYAKU Advanced Materials, Inc.). The polymer layer was formed by spin-coating. 10 μL of PMMA solution was poured onto the tapered structures on the glass substrate. First, it was spin-coated at 500 rpm for 5 s, and the rotation rate was monotonically increased from 500 to 1,000 rpm for 30 s, and continued for 60 s at 1,000 rpm. The PMMA layer was obtained with a thickness of ~100 nm, which was sufficiently thin to observe the tapered structure by SEM (SU9000, Hitachi High-Tech, Corp.), as shown in Fig. S5(b), although the contour of the structure became blurred. Here, a thin platinum layer was coated with thickness of 1.5 nm as a conductive layer on the PMMA layer by an ion sputtering device (JEC-3000FC, JEOL Ltd.) to observe it in SEM. Finally, we created a hole using FIB lithography near the apex of the tapered structure, as shown in Fig. S5(c). The FIB system allows accurate positioning of the hole with nanometric precision such that the apex remains sharp. The hole size was $500 \times 500$ nm$^2$.

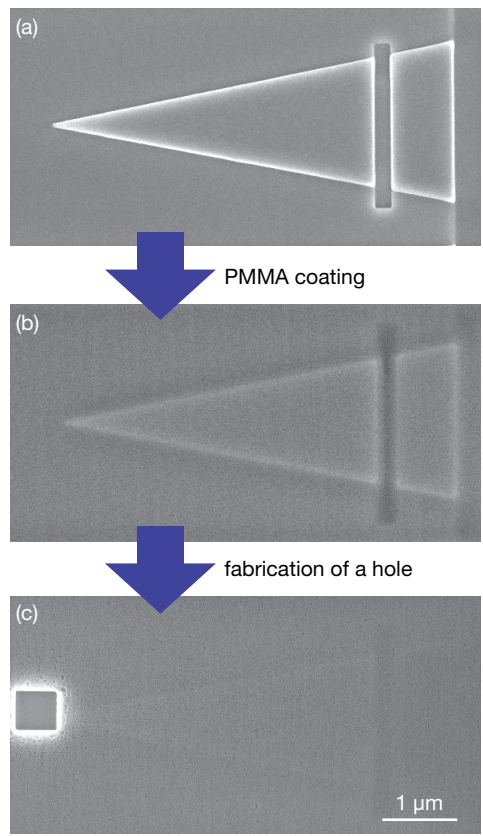

PMMA coating

fabrication of a hole

1 µm

**Figure S5.** SEM images of the aluminum tapered structure at each fabrication process. (a) Aluminum tapered structure fabricated by FIB milling. (b) Aluminum tapered structure embedded in a PMMA layer. (c) Aluminum tapered structure in the PMMA layer with a hole fabricated by FIB milling.

## 6.    Surface roughness of an aluminum tapered structure

The surface roughness of the aluminum tapered structure is crucial for facilitating efficient plasmon propagation toward the apex. Thus, we determined that the surface was sufficiently smooth. Figure S6 shows a topographic image of an aluminum layer deposited on a coverslip, which was obtained by atomic force microscopy (AFM) (MFP-3D-BIO, Oxford Instruments). The aluminum layer was prepared under the same conditions as those described in Section 5. A topographic line profile is obtained along the blue dotted line, as shown in the inset. The height fluctuated within a range of 2 nm. The root mean square (RMS) was only 0.86 nm, which was highly smooth for efficient plasmon propagation, as discussed in previous works[3].

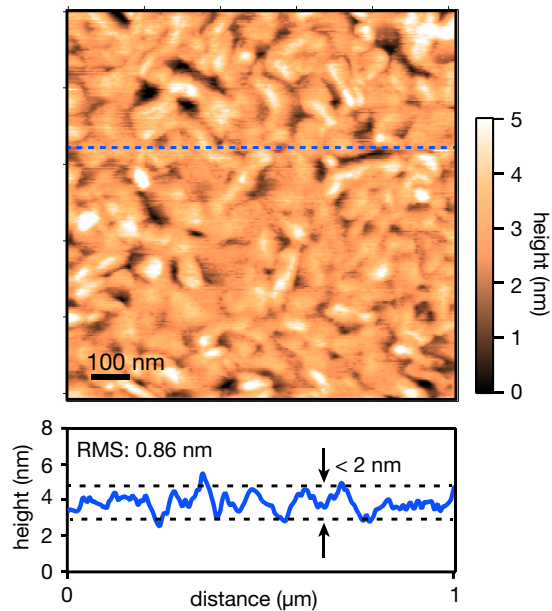

**Figure S6.** AFM image of an aluminum layer deposited by vacuum vapor deposition, and a topographic line-profile obtained along the blue dotted line.

## 7. Experimental setup and procedure for nanoscale optical switching

Figure S7 illustrates the experimental setup used for nanoscale optical switching. A supercontinuum laser (SuperK COMPACT, NKT Photonics) was used as the visible light source, and a UV laser (LSR375NL-150, NaKu Technology Co., Ltd.) was used as the UV light source. Mechanical shutters (SHB1T, Thorlabs, Inc.) were inserted in both the incident paths to control the laser exposure time. Some wave plates were located to adjust the direction of the incident polarizations perpendicular to the slit of an aluminum tapered structure such that the excitation efficiency was maximized for plasmon nanofocusing. An oil-immersion objective lens (NA: 1.45, UPLXAPO100XO, Olympus) was used to tightly focus the laser on the slit to excite the plasmon nanofocusing. We observed optical images of plasmon nanofocusing, shown in Fig. 5 in the main text, through the CMOS camera1 (VLG2020e-SP, Visualix).

We monitored the scattered signal generated by visible plasmon nanofocusing from the apex of the tapered structure to distinguish two states of diarylethene molecules: closed- and open-ring forms. A pinhole (diameter: 200 μm) was used as a confocal element to extract the scattered signals only from the apex. The pinhole was accurately positioned at the apex by observing it through the CMOS camera 2. The scattered signals were detected by a Peltier-cooled CCD (PIXIS100BR_eXcelon, Teledyne Technologies Inc.) as a scattering spectrum through a spectroscope (IsoPlane 160, Teledyne Technologies Inc.) with grating (150 g/mm). It is possible to distinguish the two states of diarylethene molecules by observing the absorbance change in the scattering spectra.

10 μL of 100 mM diarylethene solution was poured onto the substrate and dried for 15 min. To analyze the absorbance change, first, we acquired a scattered spectrum as a reference by visible plasmon nanofocusing, after completely turning diarylethene molecules to open-ring form under visible light irradiation. As shown in Fig. 5(c) in the main text, the absorption spectrum of open-ring diarylethene molecules exhibits a flat feature because it has almost no absorption in the wavelength range of 400–600 nm. We thereafter obtained a scattered spectrum of diarylethene molecules after exposing diarylethene molecules at the apex by UV plasmon nanofocusing for 10 s to convert them into the closed-ring form. The exposure time to obtain the scattered spectrum by visible plasmon nanofocusing was 1 s, which could slightly convert a small portion of the closed-ring diarylethene molecules to an open-ring form, but did not significantly affect the implementation of the optical switching experiments. The obtained scattering spectrum was converted to an absorption spectrum using the reference spectrum through the Lamber–Beer law (-$\log_{10}(I/I_{ref})$)). Here, $I$ and $I_{ref}$ denote the scattering and reference spectra, respectively. The absorption spectrum exhibited a peak at ~530 nm, as shown in Fig. 5(b) in the main text, which corresponds well with the ordinary absorption spectrum of closed-ring diarylethene molecules in Fig. 5(c) in the main text. We thereafter obtained the scattering spectrum after exposing diarylethene molecules to visible plasmon nanofocusing for 10 s to turn them into the open-ring form. The absorption spectrum was calculated in the same way and exhibited flat features, as they did not have absorption. We repeated this cycle for optical switching when we investigated the repeatability of the switching, as shown in Fig. 6. The time variation in the absorption spectra was evaluated by precisely controlling the exposure time using mechanical shutters, as shown in Fig. 5 in the main text. The incident powers of the supercontinuum and UV lasers were 1.65 and 1.5 µW at the slit, respectively. The exposure time of 10 s was sufficiently long to completely turn the open-ring form to closed-ring form, and vice versa. The acquisition time of the scattering spectrum using plasmon nanofocusing of the supercontinuum laser was as short as 1 s for all the measurements.

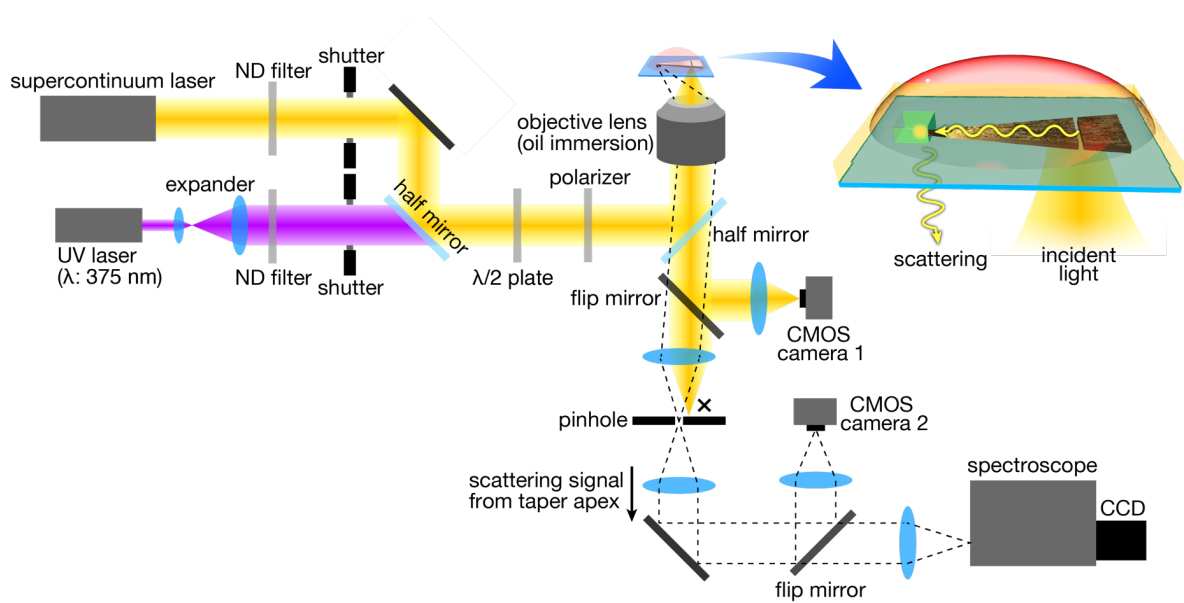

**Figure S7.** Schematic of experimental setup for the nanoscale optical switching.

## 8.  Absorption spectral change with and without a hole

To ensure that the change in the absorption spectra was attributed to the conformational change of diarylethene molecules located at the apex of the aluminum tapered structure, we investigated the absorption spectral changes with and without a hole at the apex. As seen in Fig. S8(a), when a hole was fabricated, we observed an increase in absorption by the UV plasmon nanofocusing. In contrast, when we did not fabricate a hole, the entire tapered structure was embedded in the polymer coating. Therefore, diarylethene molecules could not touch the apex when the diarylethene solution was poured onto the substrate. No absorption change should be observed in this case. Figure S8(b) shows the absorption spectra acquired before and after the UV plasmon nanofocusing without a hole. No change was observed in the absorption spectra, even after UV plasmon nanofocusing, which confirmed that the absorption change was caused by diarylethene molecules existing only near the apex of the tapered structure within a nanoscale volume.

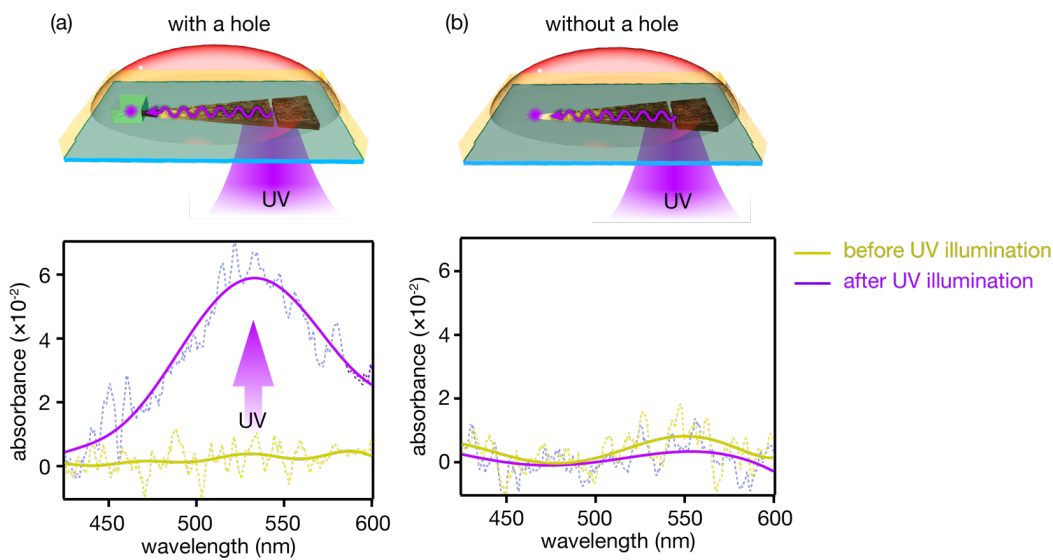

**Figure S8.** Absorption spectra obtained before and after UV plasmon nanofocusing (a) with a hole and (b) without a hole.

**9. Absorption spectra acquired through nine cycles of nanoscale optical switching**

As reported in Fig. 6(b) in the main text, we repeated the nanoscale optical switching for up to nine cycles. Figure S9 shows the absorption spectra obtained for the individual cycles. Although the spectral shapes were slightly different between the spectra, the peak positions and heights were almost the same for all cases. This confirmed that the diarylethene molecules were completely switched within 10 s, and nanoscale optical switching was highly stable and reproducible.

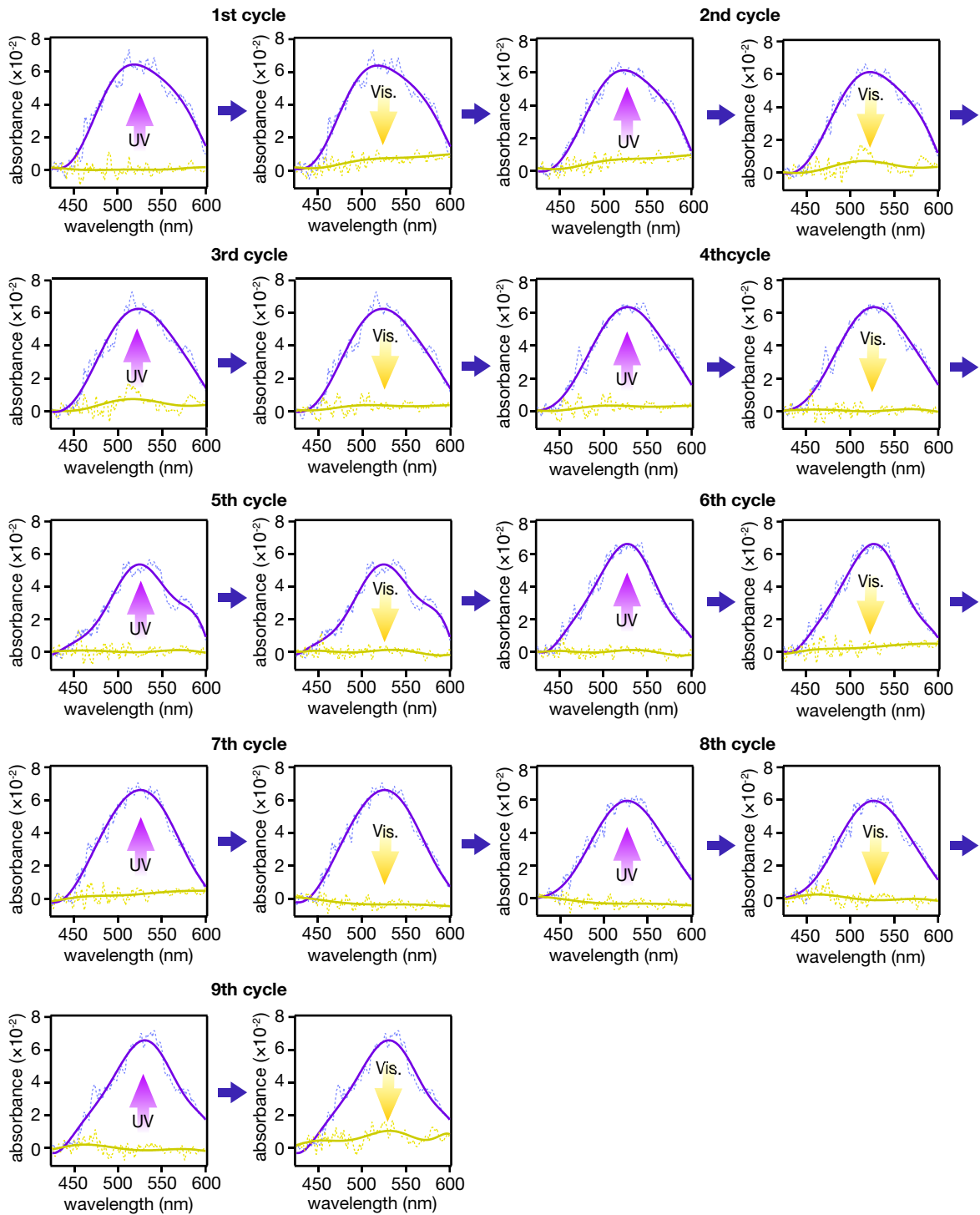

**Figure S9.** Absorption spectra acquired through nine cycles of the nanoscale optical switching